\documentclass[prb,twoside,twocolumn,showpacs,superscriptaddress]{revtex4}
\usepackage{graphicx}
\usepackage{amsmath}
\usepackage{amssymb}

\begin{document}

\title{A small sealed Ta crucible for thermal analysis of volatile metallic samples}
\author{Y. Janssen} \email[Email:]{yjanssen@ameslab.gov}
\affiliation{Condensed Matter Physics, Ames Laboratory, Ames, Iowa
50011, USA} \affiliation{Materials and Engineering Physics, Ames
Laboratory, Ames, Iowa 50011, USA}
\author{M. Angst}
\affiliation{Condensed Matter Physics, Ames Laboratory, Ames, Iowa
50011, USA}
\author {K. W. Dennis}
\affiliation{Materials and Engineering Physics, Ames Laboratory,
Ames, Iowa 50011, USA}
\author{P. C. Canfield}
\affiliation{Condensed Matter Physics, Ames Laboratory, Ames, Iowa
50011, USA} \affiliation{Department of Physics and Astronomy, Iowa
State University, Ames, Iowa 50011, USA}
\author{R. W. McCallum}
\affiliation{Materials and Engineering Physics, Ames Laboratory,
Ames, Iowa 50011, USA} \affiliation{Department of Materials
Science and Enigineering, Iowa State University, Ames, Iowa 50011,
USA}

\date{\today}
\begin{abstract}

Differential thermal analysis on metallic alloys containing
volatile elements can be highly problematic. Here we show how
measurements can be performed in commercial, small-sample,
equipment without modification. This is achieved by using a sealed
Ta crucible, easily fabricated from Ta tubing and sealed in a
standard arc furnace. The crucible performance is demonstrated by
measurements on a mixture of Mg and MgB$_2$, after heating up to
1470$^{\circ}{\rm C}$. We also show data, measured on an alloy
with composition Gd$_{40}$Mg$_{60}$, that clearly shows both the
liquidus and a peritectic, and is consistent with published phase
diagram data.

\end{abstract}


\maketitle

Differential thermal analysis (DTA) is an important tool for
determining both phase-diagram and thermal-processing information
for advanced-materials development.  It is particularly helpful in
the solution growth of single crystals~\cite{Janssen05-1}.  For a
typical commercial, small-sample, DTA, in this case a PerkinElmer
Pyris DTA 7, the standard crucible is an \emph{open} Al$_2$O$_3$
crucible. This presents two problems, one of stability and one of
volatility. To close the crucibles, Al$_2$O$_3$ caps are
available, but they do not provide sufficient sealing for samples
containing highly volatile components. In addition, elements such
as Li, Mg, and rare earths attack Al$_2$O$_3$.

For metallic alloys that do not attack it, Ta is often the
crucible material of choice. Ta is easily fabricated into
crucibles and has been found to remain inert during lengthy
high-temperature growth experiments for a wide range of metallic
alloys.  Moreover, in growth crucibles~\cite{Canfield01}, Ta caps
can be welded onto the crucibles in a standard arc furnace,
resulting in a sealed crucible with an Ar room temperature
pressure of $\sim$1 bar. For growth of crystals that involves
volatile elements, such as Mg, Li or Yb, such sealed Ta
crucibles~\cite{Canfield01} have been shown to yield well-formed
single crystals. In order to replicate the growth
environment~\cite{Janssen05-1}, small sealed Ta crucibles, usable
in a commercial DTA, are required.  Given the high melting point
of Ta, $\sim$3000$^{\circ}{\rm C}$, and the small dimensions of
the crucible, careful consideration must be given to the heat flow
during the sealing process.

In the differential thermal analyzer, Al$_2$O$_3$ crucibles are
placed in Pt sample and reference cups with Pt-based thermocouples
inserted through the bottom so that they contact the crucibles. If
Ta is placed in direct contact with the Pt, there is a possibility
that at elevated temperatures ($>$ 1000$^{\circ}{\rm C}$), Ta may
diffuse into the Pt, and thereby change the thermocouple
calibration, or it may create a diffusion bond between the Pt
parts and the Ta. Therefore, the Ta crucibles are placed inside
the standard Al$_2$O$_3$ crucibles, which can be reused multiple
times in this application.

The body of the crucible consists of small-diameter Ta tubing. In
order to successfully seal volatile metals in such a crucible, it
is necessary to keep their temperature as low as possible.
However, the radiant heat produced while sealing the crucible was
found to be sufficient to volatilize Mg. The incorporation of a
radiation shield inside of the crucible adequately addressed this
problem.

\begin{figure}[!htb]
\begin{center}
\includegraphics[width=0.4\textwidth]{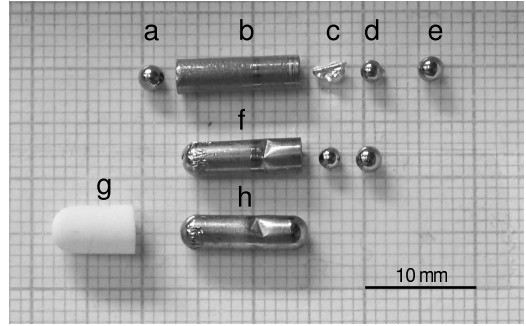}
\caption{The different steps in manufacturing the sealed
volatile-element DTA crucible, as described in the
text.}\label{Photo}
\end{center}
\end{figure}

In Fig.~\ref{Photo} the components of the crucible (a-h) are
displayed. The Ta crucibles are made of pieces of Ta tubing 10-12
mm length, 2.15 mm inner diameter and 3.2 mm outer diameter (b).
Two pieces of Ta of about 120 mg and another one of about 75 mg
are melted in a standard Ar-atmosphere arc furnace, to make Ta
balls, which form naturally upon melting, because of surface
tension. The larger two balls (a and e), are larger than the inner
diameter of the Ta tube and are used to form the ends of the
crucible, whereas the smaller ball (d), which just fits inside the
tubing, will act as a radiation shield that protects the sample
(c).

To form an \emph{unsealed} Ta crucible, the Ta tube (b) is clamped
in the water-cooled copper hearth of an Ar-atmosphere standard arc
furnace, and one of the larger Ta balls (a) is placed on top of
it. The ball is carefully melted (with a low operating current),
so that it forms a closed lid of the crucible without making it
wider than the original tube. This crucible is the standard Ta
crucible that we have used
before~\cite{Janssen05-1,Janssen05-2,Huang05}. A similar crucible
can be made by mounting the tubing in a lathe and spinning one end
closed, but this requires a skilled machinist, and can thin the
walls and leave a small hole at the tip.

Next, the sample (c) is placed inside, and a pair of cutters is
used to pinch indentations in the crucible (f). The purpose of the
indentations is to position the radiation shield (d), the smaller
Ta ball, above the sample so that the shield is in contact with
the crucible but with neither the sample nor the lid (e). The
assembly is then sealed by carefully melting the lid, the second
larger Ta ball, on top of the crucible, as above, to form the
sealed crucible (h). Finally, the standard Al$_2$O$_3$ crucible
(g) will hold the crucible in the DTA.

We have made no attempt to accurately determine the maximum sample
temperature as the crucible is sealed with the radiation shield in
place. However, we have used this technique to encapsulate several
pieces of Mg and then opened the crucible to inspect the contents.
The results indicated that partial melting of some of the Mg
pieces occurred, but the pieces were still distinct.  This
suggests that, although there was sufficient radiative heat to
bring the irregular pieces of Mg to near their melting
temperature, upon the occurrence of surface melting, the increase
in thermal contact with the crucible prevented complete melting of
the sample.  Since the vapor pressure of Mg at its melting
temperature (650$^{\circ}{\rm C}$) is only about $\sim$ 3 mbar
(Ref.~\onlinecite{Loebel69}) the increase in pressure within the
crucible is not sufficient to displace the molten ball of Ta
(Fig.~\ref{Photo}e), so that the crucible is successfully sealed.
Assuming that the average temperature inside the crucible was
about 600$^{\circ}{\rm C}$, the room-temperature Ar pressure
inside the sealed crucible was approximately 1/3 bar.

\begin{figure}[!tb]
\begin{center}
\includegraphics[trim=1.5in 0.5in 1.5in 0.5in,width=0.3\textwidth]{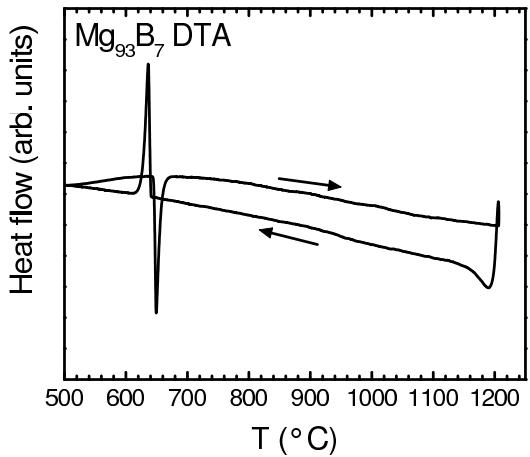}
\caption{Relevant part of the DTA curves of a sample of Mg and
MgB$_2$ measured upon heating and cooling with a 10$^\circ$C/min
rate. The melting (and solidification) event near 640$^\circ$C can
clearly be distinguished.}\label{MgBDTA}
\end{center}
\end{figure}

The effectiveness of these crucibles was tested by a series of DTA
measurements made in an attempt to determine solution-growth
parameters for growing MgB$_2$, from a mixture of Mg and MgB$_2$.
9.7 mg Mg and 0.71 mg MgB$_2$ (which amounts to
Mg$_{0.93}$B$_{0.07}$) were sealed in a crucible as described
above, and subject to several DTA runs. The highest temperature
reached was 1470$^\circ$C, which lies outside of the calibration
range of the instrument. The curve obtained from one of the runs
after the sample had been at 1470$^\circ$C, is displayed in
Fig.~\ref{MgBDTA}. The sample was heated and cooled between 450
and 1200$^\circ$C at a rate of 10$^\circ$C/min. Upon heating, only
one endothermic event, with an onset temperature of about
640$^\circ$C, was observed, and upon cooling only one exothermic
event was observed, also with an onset temperature of about
640$^\circ$C. These results are consistent with the published Mg-B
phase diagram~\cite{Okamoto00}, which shows no exposed liquidus
for MgB$_2$ at ambient pressures, and a eutectic (essentially just
the melting of pure Mg) at $\sim$650$^\circ$C. Furthermore, the
results indicated that solution-growth parameters cannot be
determined in this way.

This experiment demonstrates quite clearly the good performance of
the crucible.  The sealed crucible allowed repeated heating cycles
up to 1470$^\circ$C, and its appearance did not change visibly due
to the temperature cycles. The strength of the DTA signal due to
the melting and solidification did not visibly change in any of
the several runs that we performed, consistent with no Mg loss
during the experiments. Note that at 1470$^\circ$C, the partial
pressure of Mg is about 12 bar and the Ar partial pressure is
estimated at about 2 bar.

\begin{figure}[!tb]
\begin{center}
\includegraphics[trim=1.5in 0.5in 1.5in 0.5in,width=0.3\textwidth]{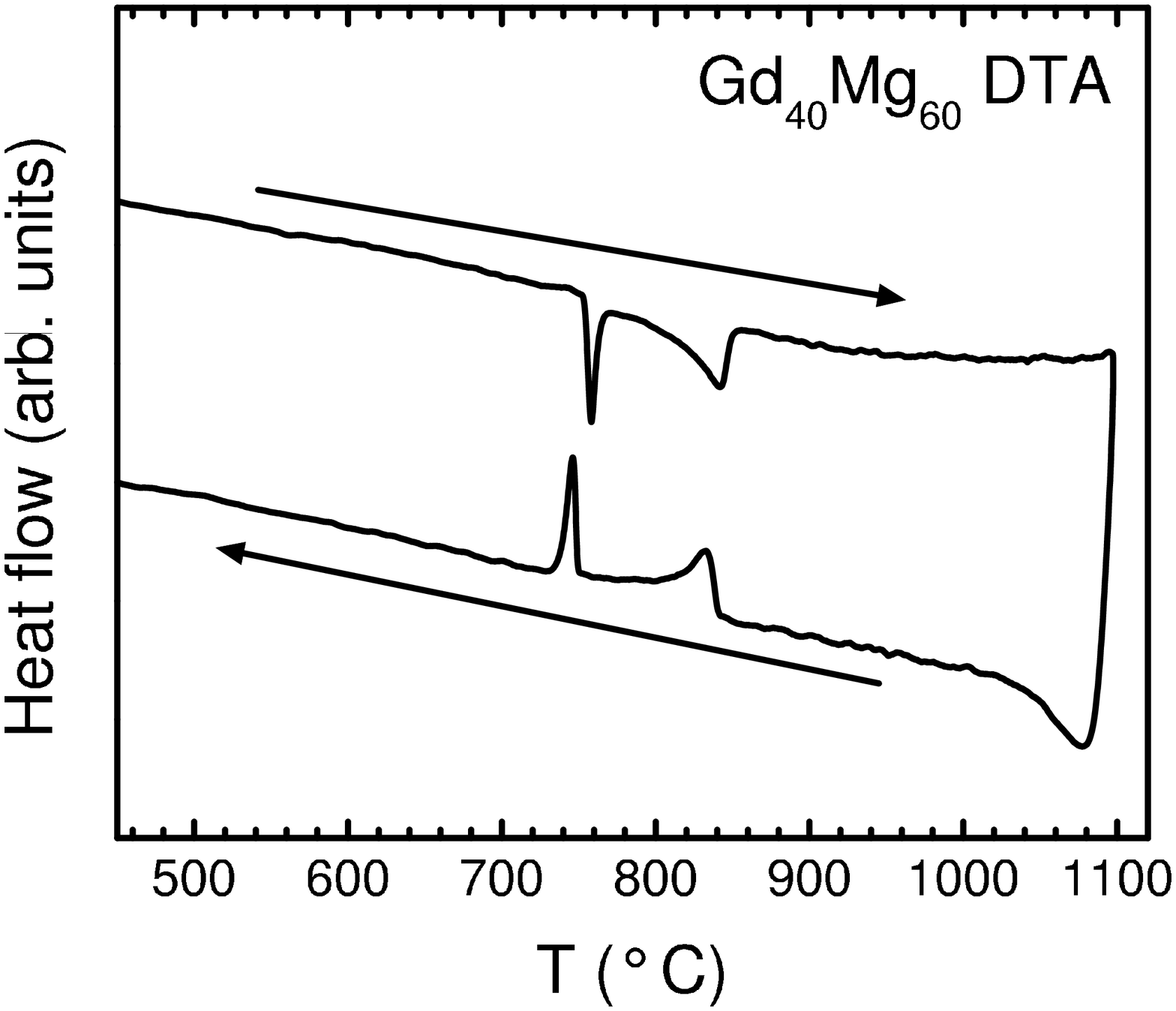}
\caption{Relevant part of the DTA curves of a sample of
Gd$_{40}$Mg$_{60}$ measured upon heating and cooling with a
10$^\circ$C/min rate. The melting (and solidication) events near
750$^\circ$C and 850$^\circ$C can clearly be
distinguished.}\label{GdMgDTA}
\end{center}
\end{figure}

A second example demonstrates the usefulness of the sealed Ta DTA
crucible for the determination of phase-diagram data. A total of
40.9 mg of Gd and Mg, corresponding to the composition
Gd$_{40}$Mg$_{60}$, was sealed in a crucible, as described above.
After alloying the elements by heating the crucible to above
1200$^\circ$C, the DTA signals were recorded. The DTA results,
measured upon heating to 1100$^\circ$C and cooling to below
400$^\circ$C at 10$^\circ$C/min are shown in Fig.~\ref{GdMgDTA}.
In both curves, two events can clearly be observed. Upon heating,
a sharp peak occurs at $\sim$750$^\circ$C, followed by an event
typical of a liquidus event~\cite{Janssen05-1}, ending near
$\sim$850$^\circ$C. Upon cooling, an event, that is also typical
of a liquidus, occurs with an onset temperature of
$\sim$840$^\circ$C, followed by a sharp peak with an onset
temperature of $\sim$750$^\circ$C. No further events were observed
at lower temperature. These results are consistent with published
phase diagram data~\cite{Okamoto00-2}, where the liquidus
temperature for the composition Gd$_{40}$Mg$_{60}$ is reported
near 850~$^\circ$C, below which the compound GdMg solidifies,
followed by the (peritectic) formation of GdMg$_2$, at
755$^\circ$C.

Both sealed and unsealed Ta crucibles fabricated as described
above have been used routinely for DTA measurements in our
laboratory for over one year. With experience, the failure rate in
sealing the crucibles is minimal. Rare failures in sealing are
easy to identify and discard and no successfully sealed crucible
has failed during measurement, so that the instrumentation has not
been exposed to reactive vapors. It should be noted that the use
of Ta crucibles requires an extremely clean environment within the
DTA, since Ta will oxidize rapidly and become brittle.  The
overall cost of the Ta crucibles is approximately the same as that
of the commercially available ceramic crucibles.

The authors are grateful to M. Huang, J. Frederick, S. A. Law, and
S. Jia for useful discussions and sample preparation. Ames
Laboratory is operated for the U.S. Department of Energy by Iowa
State University under Contract No.\ W-7405-Eng-82. This work was
supported by the Director for Energy Research, Office of Basic
Energy Sciences.

\end{document}